\documentclass[10pt,a4paper,superscriptaddress,aps,prd,showkeys,showpacs,nofootinbib,reprint]{revtex4-1}
\usepackage[utf8]{inputenc}
\usepackage{verbatim}
\usepackage{amsmath,amsfonts,amssymb}
\usepackage[breaklinks,colorlinks]{hyperref}
\usepackage{natbib}
\usepackage{tikz}
\usepackage{float}
\usepackage{graphicx}
\usepackage{epsfig}
\usepackage{adjustbox}
\usepackage{tabularx}
\usepackage{amsbsy}
\usepackage{braket}
\usepackage[singlelinecheck=off,justification=Justified,font=small]{caption}
\usepackage{subcaption}
\hypersetup{%
,urlcolor=blue
,citecolor=blue
,linkcolor=blue
}

\def\CLEAN{\sc{clean} \rm}

\begin{document}

\title{Limits on global cosmic birefringence using radio sources}
 
\author{Richard A. Battye}
\email[]{richard.battye@manchester.ac.uk}
\affiliation{%
Jodrell Bank Centre for Astrophysics, Department of Physics and Astronomy, University of Manchester, Manchester, M13 9PL, U.K.
}

\author{Neal Jackson}
\email[]{neal.jackson@manchester.ac.uk}
\affiliation{%
Jodrell Bank Centre for Astrophysics, Department of Physics and Astronomy, University of Manchester, Manchester, M13 9PL, U.K.
}

\author{Ian Browne}
\email[]{ian.browne@manchester.ac.uk}
\affiliation{%
Jodrell Bank Centre for Astrophysics, Department of Physics and Astronomy, University of Manchester, Manchester, M13 9PL, U.K.
}

\label{firstpage}

\date{\today}

\begin{abstract}
We have made measurements of the difference between the position angle (PA) on the sky and the polarization position angle (PPA) of radio sources using data from a combination of the Radio Fundamental Catalogue (RFC) across a range of frequencies between 2.7 and 15~GHz and Cosmic Lens All Sky Survey (CLASS) which observes polarization at 8.4~GHz (X-band). For the 2111 sources with jet PAs measured in the X-band and a known redshift, the distribution is peaked at $\approx 0^{\circ}$ as expected for no birefringence and it can be modelled by two populations: one which is a Gaussian with mean $\mu_{\beta}=(0.2\pm 1.0)^{\circ}$ and standard deviation $\sigma_{\beta}=(14.7\pm 1.1)^{\circ}$ and the other a uniform distribution of sources which are a fraction $f_\beta=0.72\pm 0.02$ of the total. Uncertainties in $\mu_\beta$ can be reduced to $\approx 0.6^{\circ}$ by stacking measurements of the PA from other wavebands. We find that limits of $\approx 0.1^{\circ}$ might be possible with a sample of $\sim 10^5$ similarly selected sources and that this could provide a confirmation of recent claims of global birefringence made using the Cosmic Microwave Background observations from the {\it Planck} satellite.
\end{abstract}


\maketitle

{\it Introduction: }Maxwell's equations are one of the great triumphs of 19th century physics predicting the existence and detailed properties of electromagnetic (EM) waves~\citep{Clerk1865}. The interpretation of most astrophysical observations is based on their predictions. One important feature which is predicted by Maxwell's equations is that the plane of polarization of the EM waves is perpendicular to the direction of travel and that the polarization position angle would be unchanged along a trajectory from a distant source to observation here on Earth. This assumption has been tested in a range of laboratory situations~\cite{Marsh:2015xka} but is less well tested on cosmological distance scales over which minute effects can build up into a potentially measureable effect.

Parity violating modifications to Maxwell's laws have been  suggested~\citep{Carroll1990,Carroll1998,Lue1999,Caldwell2011} which could be related to the interaction between the electromagnetic sector and a pseudo-Nambu-Goldstone (pNG) field, $\phi$, associated with a particle such as an axion. The Lagrangian for the interaction of the pNG field with the EM sector is typically of the form 
\begin{equation}
{\cal L}={1\over 2}\partial_{\mu}\phi\partial^{\mu}\phi-V(\phi)-{1\over 4}F^{\mu\nu}F_{\mu\nu}-{1\over 4}g\phi F^{\mu\nu}{\tilde F}_{\mu\nu}\,,
\end{equation}
where ${\tilde F}_{\mu\nu}=\epsilon_{\mu\nu\rho\sigma}F^{\rho\sigma}/2$ is the dual of the EM tensor $F_{\mu\nu}$, $V(\phi)$ is the pNG potential and $g$ is the coupling constant. Such models predict a rotation, known as comic birefringence (CB),  of the plane of polarization, $\beta(z)=\textstyle{1\over 2}g[\phi(0)-\phi(z)]$ where $\phi(z)$ is the spatially homogeneous value of $\phi$ at redshift $z$. If the distribution of the pNG field is spatially inhomogeneous, this can have a non-zero two-point correlation function.

Constraints on the rotation of the plane of polarization can be obtained from observations of the polarization of the Cosmic Microwave Background (CMB). Such observations can be used to estimate the so-called E- and B-mode power spectra~\citep{Kamionkowski1997,Kamionkowski1997a,Zaldarriaga1997}.  Scalar density waves only create E-modes and the detection of B-modes would possibly indicate the existence of primordial gravitational waves, vorticity, or even some kind of anisotropy of the Universe. Even then, the E- and B-mode spectra would be expected to be uncorrelated, and the cross-power spectrum, $C_\ell^{EB}$, can be used to constrain $\beta$. Progressively tighter limits on the global component of $\beta$ have been published~\citep{PhysRevLett.102.161302,WMAP:2012nax} and the 68\% confidence level (CL) range obtained using the data from the {\it Planck} Satellite is $\beta(z_{\rm LSS})=(0.21\pm 0.05 [{\rm stat}]\pm 0.28 [{\rm sys}])^{\circ}$ with the systematic uncertainty being dominated by the uncertainty in the calibration of the instrumental contribution to the polarization position angle (PPA). \citet{Minami:2020odp} have performed a self-calibration analysis of these data claiming a detection $\beta(z_{\rm LSS})=(0.35\pm 0.14)^{\circ}$ with $|\beta(z_{\rm LSS})|>0$ at more than 99\% CL. These conclusions have been supported by a range of other analyses~\citep{Diego-Palazuelos:2022dsq,Eskilt:2022cff,Diego-Palazuelos:2022cnh,Eskilt:2023nxm,Cosmoglobe:2023pgf}. Given the difficulties involved in making these measurements and also the potentially profound implications, it seems reasonable to search for alternative ways of constraining, or measuring CB.

The PPA of a galaxy, $\alpha=\textstyle{1\over 2}\tan^{-1}(U/Q)$ (where $Q$ and $U$ are the Stokes parameters), can be used to provide an estimate of $\beta$. This can be done by assuming that some aspect of the measured shape of the source is related to the PPA; something which can be expected on physical grounds for highly polarized sources~\citep{Carroll1990,Harari:1992ea}. Measuring $\alpha$ and the structural position angle (PA), $\theta$, then $\beta=\theta-\alpha-90^{\circ}$ taking into account that $\beta\in(-90,90)^{\circ}$\footnote{We measure the PA and PPA using the IAU convention of North through East, which is the opposite definition used in CMB observations}. Of particular relevance here are galaxies with Active Galactic Nuclei (AGN) whose emission is often dominated by  highly directional jets and which can be observed in a number of wavebands - a summary of some observations of individual galaxies is presented in \citet{Kaufman:2014rpa} with differences between $\alpha$ and $\theta$  being typically $\sim$ a few degrees. However, these sources have been ``cherry-picked'' to be very well aligned which could be a coincidence and moreover there are limits on how well the angles can be measured. In this {\it letter} we have used a large sample of radio sources to constrain the distribution of $\beta$ using a statistical approach. The sample is not necessarily selected in an unbiased way, but importantly it has also not been chosen to have low values of $\beta$. 

\begin{figure}[t]
\begin{center}
\resizebox{0.45\textwidth}{!}{\includegraphics{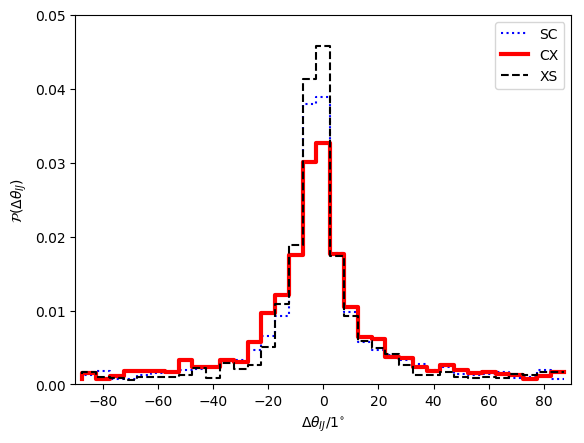}}
\end{center}
\caption{Probability distribution of $\Delta\Theta_{IJ}$ where $I,J=\{S,C,X\}$. We have not included U because there are relatively few sources.}
\label{fig:comp}
\end{figure}

{\it Data: } As explained in the supplementary material~\cite{Supp} we have assembled a sample of 6448 radio sources with a measurements of a PA, $\theta_I$ for at least one of $I=\{S,C,X\}$, using the Radio Fundamental Catalogue (RFC)~\cite{2025ApJS..276...38P} and PPA, $\alpha$, from the Cosmic Lens All Sky Survey (CLASS)~\cite{Jackson:2007vm}. In addition this includes refs.~\cite{1974A&AS...15..417H,2006myerstaylor,2023ApJ...944..107C,2023AJ....165..124A,2023OJAp....6E..49F,2025ApJS..276...38P}. In Fig.~\ref{fig:comp} we present a comparison between the measurements of $\theta_I$  by plotting the probability distribution $\Delta\theta_{IJ}=\theta_I-\theta_J$ taking into account wrap-around effects. Although there is some dispersion between the different frequencies, probably due to them probing different parts of the source, this clearly indicates that the PAs are consistently measured. We have also compared the position angles to those measured in ref.~\cite{Joshi:2007yf} which were computed by visual inspection of individual images and there is very good agreement. Focusing specifically on $\Delta\theta_{CX}$ we find that the distribution is compatible with a dispersion of $\approx 10^{\circ}$ and a failure fraction, modelled by a uniform distribution, of $\approx 0.25$. The polarization signal-to-noise ratio is defined by ${\rm SNR}=(Q^2+U^2)/\sqrt{Q^2\sigma_Q^2+U^2\sigma_U^2}$ where $\sigma_I$ for $I=\{Q,U\}$ is the measured noise, typically $\approx 0.3\,{\rm mJy}$, for each source. If the ${\rm SNR}>3$ then the r.m.s. uncertainty in the PPA $\langle\Delta\alpha^2\rangle^{1/2}\approx 10^{\circ}$ and hence we expect an overall dispersion in $\beta$, ignoring possible calibration uncertainties, by adding the two uncertainties in quadrature to give $\sim 14^{\circ}$. We are unable to quantify possible biases in the estimates of $\beta$ derived from this procedure, but we do not claim a detection using this sample. Future work will be necessary to develop unbiased estimators and precisely quantify the uncertainties in the calibration, before this technique can be applied to larger samples - see the later discussion. We note that we do not expect the effects of Galactic Faraday Rotation (FR) to be significant at X-band since typical Galactic ${\rm RMs}\sim 10\,{\rm rad}\,{\rm m}^{-2}$ and they will not be coherent across the sky; the effects of FR internal to the sources could add some dispersion to the measurement. Again this is unlikely to be coherent. We have investigated the effects of made corrections to the PPAs using a Galactic Faraday rotation map~\cite{2022A&A...657A..43H} and found that correction to the inferred limits on $\beta$ were $<0.1^\circ$ backing up our assertion above.

\begin{figure}[t]
\begin{center}
\resizebox{0.45\textwidth}{!}{\includegraphics{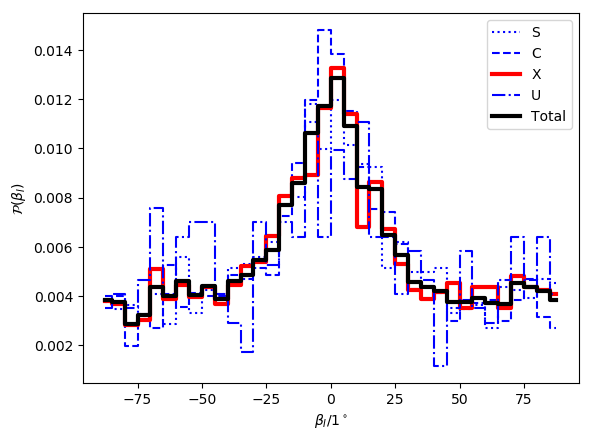}}
\end{center}
\caption{Probability distribution ${\cal P}(\beta_I)$ for $I=\{S,C,X,U\}$ and total for the subset defined in column 3 of Table~\ref{tab:constraints}, that is, with ${\rm SNR}>3$ and where there is a known redshift. The cases of X-band and `Total' are highlighted since they are emphasized in the discussion.}
\label{fig:beta}
\end{figure}

 We have used the data to estimate the probability distribution ${\cal P}(\beta)$ - calculated by creating a histogram with 36 bins and normalizing such that $\int{\cal P}(\beta)d\beta=1$ - and these results are presented in Fig.~\ref{fig:beta}. We see that all the distributions are peaked around $\beta\approx 0^{\circ}$, but that they do not fall-off to zero at $\pm 90^{\circ}$. U-band is the most noisy due to the relatively small number of sources, but all the bands look consistent with each other. In what follows we will concentrate on X-band, which has the largest number of sources and has both $\theta$ and $\alpha$ measured at the same frequency, and `total' which includes measurements at all frequencies\footnote{In what follows we will treat the measurements at the different bands as independent measurements and ignore the fact that $\alpha$ is only measured once. We will see that the dominant source of uncertainty on $\beta$ comes from a uniform background and hence the covariance introduced by using a single measurement of $\alpha$ should be subdominant.}.

{\it Modelling the distribution of $\beta$ for radio sources: }
One can attempt to model the measured distribution of $\beta$ from the average and standard deviation of the distributions presented in Fig.~\ref{fig:beta}, but this is fraught with dangers of bias since it can become dominated by large asymmetric fluctuations in the uniform background. We have derived limits by modelling the probability distribution ${\cal P}(\beta)$ comprising two components: one is a truncated Gaussian distribution with mean $\mu_\beta$ and standard deviation $\sigma_{\beta}$ and the other is a uniform background which make up a fraction $f_\beta$ of the sources. We then estimate the likelihood of the model parameters $(\mu_\beta,\sigma_\beta,f_\beta)$ with respect to the data and marginalize over the other two to create a probability distribution for each of the parameters to allow conclusions to be drawn. We have tested our analysis pipeline by including offsets in $Q$ and $U$,  added to each of the sources, that might come from calibration uncertainties and also a spurious constant offset in the values of $\alpha$ which was easily detected when larger than the statistic uncertainties found using the real data.

\begin{table}
    \centering
    \small
    \setlength{\tabcolsep}{5pt}  
    \renewcommand{\arraystretch}{1.2}  
    \begin{tabular}{c|c|c|c|c}
        Selection & All&SNR$>3$&$z$&$|\Delta\theta_{\rm CX}|<20^{\circ}$ \\
        \hline

		$\theta_{\rm S}$ & $0.7\pm 1.5$ & $1.0\pm 1.4$ & $0.7\pm 1.5$ & $0.2\pm 2.1$ \\
        $\theta_{\rm C}$ & $1.1\pm 1.2$ & $1.3\pm 1.0$ & $0.9\pm 1.0$ & $0.1 \pm 1.1$ \\
        $\theta_{\rm X}$ & $1.3\pm 1.3$ & $0.3\pm 1.0$ & $0.2\pm 1.0$ & $-0.4\pm 1.1$ \\
        $\theta_{\rm U}$ & $1.5\pm 4.6$ & $2.0\pm 4.0$ & $1.8\pm 3.9$ & $-5.2\pm 4.6$ \\
		\hline
        Total& $1.1\pm 0.7$ & $0.9 \pm 0.6$ & $0.6\pm 0.6$ & $-0.3\pm 0.7$ \\
	\end{tabular}
    \caption{Constraints on $\mu_{\beta}/1^{\circ}$.Uncertainties are defined at the 68\% CL. $z$ denotes that the catalogue contains a known redshift. The row `all' refers to all sources in the catalogue and `total' to adding together all wavebands. The column NR means no restriction. From the left the subsets include all constraints from the previous column. The number of sources in each subset is shown in the supplementary material.}
    \label{tab:constraints}
\end{table} 

\begin{figure}
\begin{center}
\resizebox{0.45\textwidth}{!}{\includegraphics{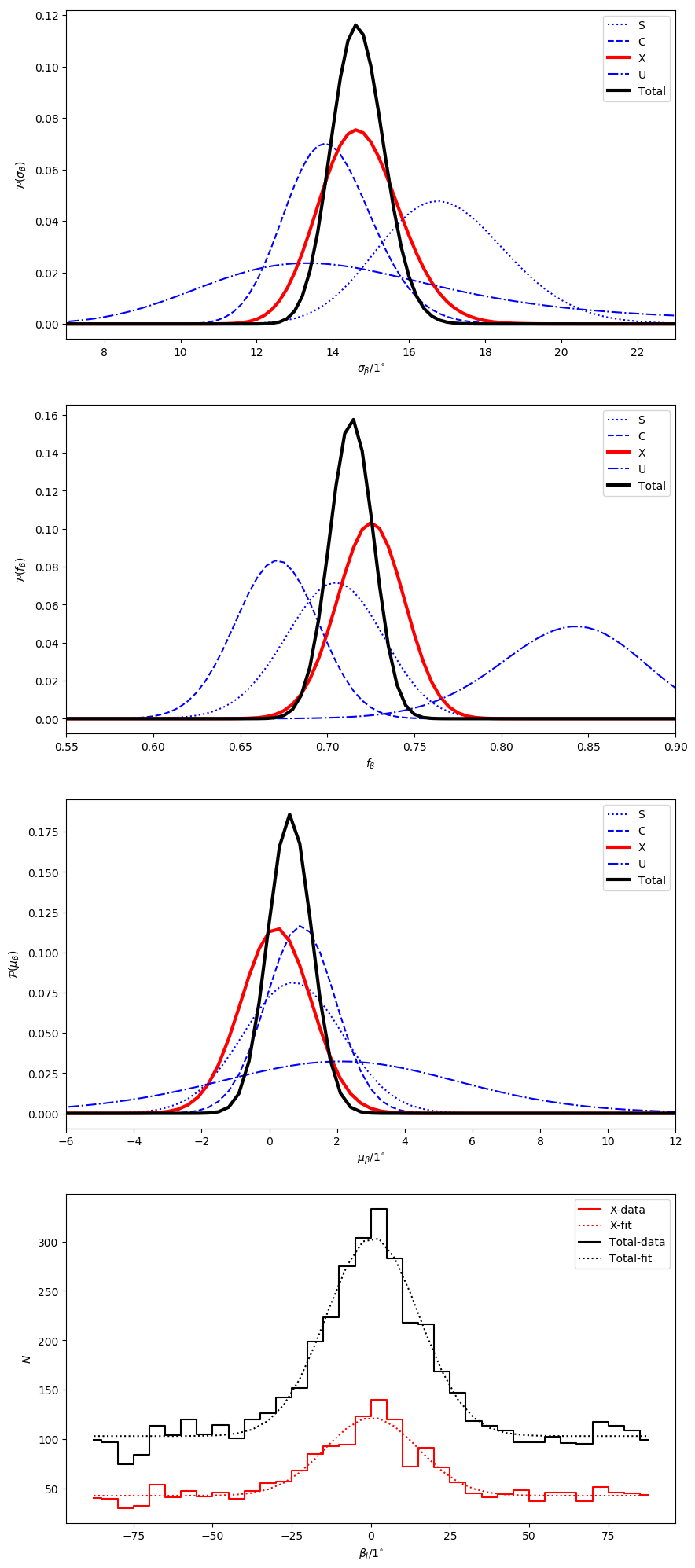}}
\end{center}

\caption{The marginalized likelihoods for the fitting parameters for the probability distributions in column 3 of table~\ref{tab:constraints} for $\theta_X$ and `total'. The top left is $\mu_{\beta}$, top right $\sigma_{\beta}$ and bottom left $f_\beta$. The bottom right shows the maximum likelihood model compared to the data presented in figure~\ref{fig:beta}.}
\label{fig:fit}
\end{figure}

For completeness, in Table~\ref{tab:constraints} we present the estimates of $\mu_\beta$ for various subsets; all are compatible with zero. The uncertainties vary very little, with decreases in the number of sources as one goes left to right in the table being offset by a cleaner sample - lower values of $f_\beta$ and $\sigma_\beta$. We will concentrate on the case in column 3, which has ${\rm SNR}>3$ and a known redshift, leading to the lowest uncertainty. The marginalized posterior distributions are presented in Fig.~\ref{fig:fit}. The values of $\sigma_{\beta}\approx 15^{\circ}$, which is compatible with the measurement error of $\approx 10^{\circ}$ in both PA and PPA, and $f_{\beta}\approx 0.7$, 0.25 of which one might attribute to the imperfections in measuring PA (see Fig.~\ref{fig:comp}) and the rest to sources where, probably for a variety of reasons, there is no correlation between PA and PPA. We note that if we make an `opposite' assumption, that is, there is no known redshift, then we get an almost flat distribution for $\beta$ indicating that there could be two populations in the sample. When we combine all measurements of $\beta$, assuming as we have argued earlier that these are independent, we find the best possible uncertainty of $\beta=(0.6\pm 0.6)^{\circ}$. We caveat our conclusions with the possibility that a global mismatch of the PPA calibration could mask a non-zero cosmological signal which we consider to be very unlikely.

{\it Discussion:} Our current best uncertainty in $\beta$ is a factor of $\sim 4-6$ greater uncertainty than the current CMB constraints dependent on which result one compares to, but is better than published constraints from individual sources~\citep{Leahy:1997wj,Wardle:1997gu,Carroll1998}. Here, we present the study of radio sources as a way of confirming, or refuting, the claimed detection from the CMB. We note that measurements from radio sources and the CMB are not directly comparable and when comparing to a specific model for $\beta(z)$ one would need to convolve that model with the redshift distribution of the sources. It is clear that the approach could be extended to much larger samples than the present one. Such approaches would need to be accompanied by more careful calibration of polarization systematics than were possible using the CLASS observations that were not originally designed for this purpose. In particular, contemporaneous calibration of absolute position angle of polarized calibration sources \cite{2013ApJS..206...16P} is needed in order to reduce the systematic error in absolute position angle. In addition, higher signal-to-noise observations of calibrators used to correct instrumental polarization would also be important in reducing errors of similar size in the derived polarization angles.

There are $\sim 10^{6}$ jet-dominated sources in the Universe with flux densities $\gtrsim 1\,{\rm mJy}$ and it might be expected that, at some point the future, these sources will be mapped to a similar level of noise as the RFC using the Square Kilometre Array (SKA). Modelling the distribution of $N$ such sources as a {\em mixture} of two populations with a fraction $f_\beta$ having $\beta\sim U(-a,a)$ and $1-f_\beta$ being $\sim N(0,\sigma_\beta^2)$ truncated at $-a$ and $a$ where $a=90^{\circ}$, we can deduce that the variance of the mixture is 
\begin{equation}
\sigma^2={1\over 3}f_\beta a^2+(1-f_\beta)\sigma_\beta^2{I_2\left({a\over\sigma_\beta}\right)\over I_0\left({a\over\sigma_\beta}\right)}\,,
\end{equation}
where $I_n(y)=\int_{-y}^y x^ne^{-{1\over 2}x^2}dx$ and hence $I_2(y)=I_0(y)-2ye^{-{1\over 2}y^2}$. In our case $a/\sigma_\beta\approx 7$ and hence it is a good approximation that $I_2/I_0\approx 1$. For $f_\beta\approx 0.7$ this implies that $\sigma\approx 90^{\circ}\sqrt{f_\beta/3}\approx 45^{\circ}$ and the measurement error on $\mu_\beta$ will be $45^{\circ}/\sqrt{N}$. Therefore, we find that uncertainties of $0.1^{\circ}$ might be possible with $N\sim 10^5$ sources. We note that this is the variance of the distribution, not what one would get from the marginalization of the likelihood. We have tested this argument with simulations and, for the relevant values of $f_\beta$ and $\sigma_\beta$, it is a good approximation.

A clear issue in the above analysis is the relatively large value of $f_\beta\approx 0.7$ that we found; that is, increasing $N$ is used to mitigate the effects by reducing the impact of the uniform background rather than random errors on the measurement of $\mu_\beta$. The value of $f_\beta$ may be intrinsic to any such observations, but could be related to our choice of data and measurement methods. The CLASS data - which is chosen since the data already exist - might not optimal for these purposes and a more bespoke polarization survey might lead to a reduction in $f_\beta$. However, ultimately it will be that the uniform distribution will come to dominate the uncertainties. Ideally, one would have some additional observations which would allow one to select objects ab-initio to have an alignment. If this could be done for $\sim 10^4$ objects, uncertainties $\sim 0.1^{\circ}$ would be possible. We have spent some time looking at the sample and, at this stage, we have not found any obvious way to reject the sources that form the uniform distribution. Obviously to reduce the uncertainties  to $\sim 0.1^{\circ}$ would require very accurate calibration of the absolute PPAs to a similar level of accuracy which might be achieved using further observations of Mars~\cite{Perley}. By only including sources with polarized fraction $\sqrt{Q^2+U^2}/I>0.02$, for which we expect a stronger correlation, we found that $f_\beta\approx 0.5$, but this reduces the number of sources such that the overall uncertainty on $\mu_\beta$ is slightly increased. This remains a challenge for the future, but the benefits would be very significant.

We should comment that there are two other approaches that have been suggested to perform similar analyses using radio sources selected in other ways. The first~\citep{Whittaker:2017hnz} makes the assumption that the iso-contours of flux density in large radio sources are statistically correlated with the direction of the position angle of polarized flux measured in the same direction. This technique, which was shown to lead to a constraint of $\beta=(-2.03\pm 0.75)^\circ$ using data from $\sim 30$ sources, builds on earlier ideas put forward by \citet{1991ApJ...367L...1K,1996ApJ...472..115K}. The number of sources to which this technique can be applied is limited by the number of suitable large radio sources. We have not done a detailed estimate, but numbers $\sim 10^5-10^6$ might be possible with the SKA dependent on resolution.

An alternative is to use ordinary star-forming galaxies~\citep{Yin:2024fez,Dai:2026zaj} together with a simple extension of ideas that have been proposed to correct radio weak lensing surveys for the effects of intrinsic alignment~\citep{Brown:2010rr,Brown:2011db,Whittaker:2015fma} relying on a different correlation between the shape of the galaxy and the polarization position angle~\citep{Zhou:2025bvz} . Large scale surveys using the SKA phase I~\citep{SKA:2018ckk} might find $\sim 10^8$ such galaxies. Depending on how many can be resolved, in order to calculate the shape of the galaxy, this could become a powerful technique.

Finally, we note that \citet{Naokawa:2025shr} has recently shown that a particular class of slowly-rolling scalar field coupled to EM using the type of coupling discussed earlier could lead to a very specific functional dependence as a function of redshift. This strongly motivates a similar study to ours with bins of $\beta(z)$. It seems clear that measuring birefringence using radio sources will continue to be an active area of research for some time to come.

\vskip 0.5cm
\centerline{\bf Supplementary Material}

The data we have used are from two sets of observations: $\theta_I$ for $I=\{S,C,X,U\}$ corresponding to observations at frequencies 2.7, 5, 8.4 and $15\,{\rm GHz}$, respectively is measured using the Radio Fundamental Catalogue (RFC) \cite{2025ApJS..276...38P}, while $\alpha$ comes from the Cosmic Lens All-Sky Survey (CLASS)~\cite{Jackson:2007vm}.  The RFC data were taken at a range of frequencies between 2.7 and 15~GHz that have typical resolutions between 0.5 and 3 milli-arcsecond (mas), corresponding to physical scales $\sim 2.5-15\,{\rm pc}$ at a typical distance of $1~{\rm Gpc}$. In contrast, the CLASS data were taken using the Very Large Array (VLA) A-configuration at 8.4~GHz (X-band) with a resolution of $\sim 200\,{\rm mas}$ which corresponds to physical scales $\sim 1\,{\rm kpc}$ at the same distance. Therefore, the polarized emission is from a much larger region of space (up to $\times$400 bigger) than that to which the RFC is sensitive. The ratio of the RFC measured flux - typically, but not always at X-band - and that from CLASS is an indicator that there is extended emission not picked up in the RFC maps or that the sources have varied between observations. We find that this ratio is peaked at around 0.8. Some of the departure from unity could arise from variability as indicated by there being  values $\geq1$ which can only arise from variability. But overall the distribution tells us that there is little extended emission missed in the RFC maps and that it is likely that the CLASS measured polarization is dominated by the parts of the source observed in RFC.  We have nevertheless repeated our analyses using sources where the ratio is $>0.6$ and this has no impact on our results. In any case, the bending of the jet between the two scales probed by the observations would be equally likely to be in either direction and, hence, increase the dispersion without a bias.

The absolute PPA calibration of CLASS data originally assumed a PPA of $-70^{\circ}$ and $33^{\circ}$ for the calibrator sources 3C48 and 3C286, respectively \citep{2006myerstaylor}. However, subsequent work \citep{2013ApJS..206...16P} showed variability in the PPA of these calibrators with time, together with a dependence on frequency. We have corrected the PPAs \cite{Jackson:2007vm} to the frequency and observation time used in CLASS; this has been done by adding 0.5$^{\circ}$ for data originally calibrated using 3C286, and $4^{\circ}$, $5^{\circ}$ and 6$^{\circ}$ for the 1994, 1995 and 1998 data originally calibrated using 3C48.

\vskip 0.2cm

\begin{figure}[h!]
\begin{center}
\resizebox{0.45\textwidth}{!}{\includegraphics{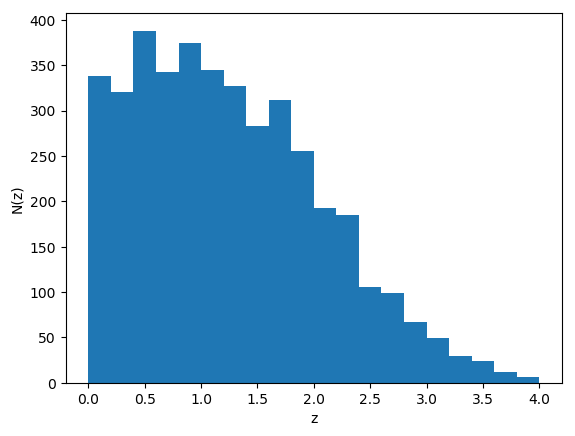}}
\end{center}
\vskip -0.8cm
\caption{Distribution of redshifts for all sources in the catalogue.}
\label{fig:Redshift}
\end{figure}

\vskip -0.2cm

\begin{table}[h]
    \centering
    \small
    \setlength{\tabcolsep}{5pt}  
    \renewcommand{\arraystretch}{1.2}  
    \begin{tabular}{c|c|c|c|c}
        Selection & NR &SNR$>3$&$z$&$|\Delta\theta_{\rm CX}|<20^{\circ}$ \\
        \hline
        All&6448&3640&2646&917\\
        \hline
		$\theta_{\rm S}$&2043&1623&1323&573\\
        $\theta_{\rm C}$&3099&1857&1405&917\\
        $\theta_{\rm X}$&4470&2807&2111&917\\
        $\theta_{\rm U}$&400&375&343&196\\
		\hline
        Total&10012&6662&5182&2603\\
	\end{tabular}
    \caption{Number of sources in subsets used here. $z$ denotes that the catalogue contains a known redshift. The row `all' refers to all sources in the catalogue and `total' to adding together all wavebands. The column NR means no restriction. From the left the subsets include all constraints from the previous column.}
    \label{tab:numbers}
\end{table}

An automatic method to estimate $\theta_I$ for each of the RFC observations has been developed. This works by examining the \CLEAN components \cite{1974A&AS...15..417H} attached to each of the uniformly-generated images \cite{2025ApJS..276...38P}. \CLEAN components were rejected if they are less than the full width at half maximum (FWHM) of the \CLEAN beam from the centre, taking into account the ellipticity of the beam. Components were also rejected if their flux density was less than 3.5 times the rms noise in the image, or less than 7 times the rms noise in the case of \CLEAN components $>20$\,mas from the centre of the source. The PA was then calculated as a flux-weighted average of the remaining position angles. Referencing everything with respect to the centre  of the source, this was done by taking an average of a set of complex numbers $F_j{\rm e}^{2i\phi_j}$, where $F_j$ and $\phi_j$ are the flux and position angle of the $j$th \CLEAN component, and using half the argument of the resulting complex number. Outliers of more than 2.8 times the median deviation from this average angle were rejected, and the angle finally recalculated using the remaining \CLEAN components. We checked 100 of the resulting angles with the RFC maps by eye to verify that the results are reasonable. Of the 21942 RFC sources, 12529 are coincident with CLASS sources, of which 6448 have a measured $\theta_I$ at at least one frequency. The numbers in various subsets of the sample are listed in Table~\ref{tab:numbers}. Redshifts for the sources were found using the {\it milliquas} survey \cite{2023OJAp....6E..49F} and, where these were not available, in the DESI quasar survey \cite{2023ApJ...944..107C,2023AJ....165..124A} and their redshift distribution is presented in fig.~\ref{fig:Redshift} which shows that they all have $z<4$ with the number in each bin increasing to a constant number somewhere in the range $1.5<z<2.0$. This distribution is a consequence of the selection functions of the parent surveys. These yield redshifts for around 2/3 of the sample, nearly all of which are likely to be quasars. Histograms of both the PPAs and PAs are individually consistent with a uniform distribution.

\bibliographystyle{apsrev4-1}
\bibliography{refs.bib}

\end{document}